\documentclass[9pt,twocolumn,twoside]{osajnl}
\journal{ol} % Choose journal (ao, aop, josaa, josab, ol, pr)

\setboolean{shortarticle}{true}
\title{Frequency-scanned microresonator soliton comb with the tracking of the frequency of all comb modes}

\author[1,2,*]{Naoya Kuse}
\author[3]{Gabriele Navickaite}
\author[3]{Michael Geiselmann}
\author[1,4]{Takeshi Yasui}
\author[1,5]{Kaoru Minoshima}

\affil[1]{Institute of Post-LED Photonics, Tokushima University, 2-1, Minami-Josanjima, Tokushima, Tokushima 770-8506, Japan}
\affil[2]{PRESTO, Japan Science and Technology Agency, 4-1-8 Honcho, Kawaguchi, Saitama, 332-0012, Japan}
\affil[3]{LIGENTEC SA, EPFL Innovation Park L, Chemin de la Dent-d'Oche 1B, Switzerland  CH-1024 Ecublens, Switzerland}
\affil[4]{Graduate School of Technology, Industrial and Social Sciences, Tokushima University, 2-1, Minami-Josanjima, Tokushima, Tokushima 770-8506, Japan}
\affil[5]{Graduate School of Informatics and Engineering, The University of Electro-Communications, 1-5-1 Chofugaoka, Chofu, Tokyo 182-8585, Japan}
\affil[*]{Corresponding author: kuse.naoya@tokushima-u.ac.jp}

\begin{abstract}
Rapid and large scanning of a dissipative Kerr-microresonator soliton comb with the characterization of all comb modes along with the separation of the comb modes is imperative for the emerging applications of the frequency-scanned soliton combs. However, the scan speed is limited by the gain of feedback systems and the measurement of the frequency shift of all comb modes has not been demonstrated. To overcome the limitation of the feedback, we incorporate the feedback with the feedforward. With the additional gain of > 40 dB by a feedforward signal, a dissipative Kerr-microresonator soliton comb is scanned by 70 GHz in 500 $\mu$s, 50 GHz in 125 $\mu$s, and 25 GHz in 50 $\mu$s (= 500 THz/s). Furthermore, we propose and demonstrate a method to measure the frequency shift of all comb modes, in which an imbalanced Mach-Zehnder interferometer with two outputs with different wavelengths is used. Because of the two degrees of freedom of optical frequency combs, the measurement at the two different wavelengths enables the estimation of the frequency shift of all comb modes. 
\end{abstract}

\setboolean{displaycopyright}{true}

\begin{document}

\maketitle

Dissipative Kerr-microresonator soliton combs (hereafter called soliton combs) have been generated from CMOS compatible platforms, providing the potential to be fully integrated optical frequency combs \cite{Herr_soliton, kippenberg2018dissipative}. After various demonstrations such as ultra-fast ranging \cite{trocha2018ultrafast, Vahala_distance18}, optical frequency synthesizers \cite{Papp_synthesizer18}, dual-comb spectroscopy \cite{suh2016microresonator, yu2018silicon}, and low phase noise microwave generation \cite{liang2015high, liu2020photonic} using free-running or frequency-stabilized soliton combs, recently, frequency-scanned soliton comb explores new type of applications. One distinguished example is ultra-fast frequency-modulated continuous-wave (FMCW) LiDAR \cite{riemensberger2020massively}, in which a frequency-scanned soliton comb is used as a set of massive parallel frequency-scanned cw lasers. Another example is scanning soliton comb spectroscopy \cite{yu2017microresonator,kuse2020continuous,lin2020broadband}. The frequency-scanned soliton comb overcomes the sparsity of the soliton comb, which limits the frequency resolution for spectroscopy. To fully open up the potential of the frequency scanned soliton comb such as high-speed and precise measurements, two fundamentals have to be developed, that is (1) rapid and large scanning and (2) measurement of the frequency shift of all comb modes. 

When soliton combs are scanned, both the frequencies of a pump cw laser and a resonance of a microresonator have to be scanned simultaneously to keep the detuning between the pump cw laser and the resonance in a range where soliton combs are maintained. Conventionally, the detuning is fixed by employing a feedback loop based on Pound-Drever-Hall (PDH) locking \cite{kuse2020continuous, Papp_PRL18} or soliton-power locking \cite{Vahala_power_mod16,lin2020broadband}. In ref \cite{kuse2020continuous} and ref \cite{lin2020broadband}, the comb mode scanning of 190 GHz (scan speed is not shown) and 31 GHz in 50 ms (= 0.62 THz/s) has been demonstrated, respectively. However, in the methods with feedback loops, the scan speed is limited by the feedback gain and dynamic range of a modulator (e.g. pump cw laser in ref \cite{kuse2020continuous, lin2020broadband}) at the modulation frequency. On the other hand, the feedforward scheme, where the frequencies of a pump cw laser and resonance are controlled by a pre-determined signal, is free from the feedback gain limit \cite{liu2020monolithic}. However, due to the non-fixed detuning, a large range scanning is very difficult without sophisticated feedforward signals. Regarding the frequency measurement, none has demonstrated the measurement of all comb modes. When a soliton comb is used for massively parallel FMCW LiDAR \cite{riemensberger2020massively}, the frequency of the soliton comb is partially measured by measuring the frequency of a beatnote between one of the comb modes and a cw laser. In ref \cite{kuse2020continuous}, where comb-resolved high-resolution spectroscopy has been demonstrated, the frequency of only one comb mode used for spectroscopy is measured by using an imbalanced Mach-Zehnder interferometer (i-MZI). In these cases, the full characterization of all comb modes cannot be achieved. 

In this letter, we demonstrate the rapid and large scanning of a soliton comb with the measurement of the frequency shift of all comb modes, showing the frequency scanning of 25 GHz in 50 $\mu$s (= 500 THz/s) with the estimated frequency error of below 10 MHz. 
The rapid and large scanning of the comb mode is enabled by incorporating the feedback with the feedforward. In addition to the feedback gain of 40 dB, the feedforward signal provides 40 dB gain when the modulation frequency of 10 kHz is applied to a microheater. Because of the large gain at the modulation frequency, the comb modes are successfully scanned by 25 GHz in 50 $\mu$s (70 GHz in 500 $\mu$s, 50 GHz in 125 $\mu$s when modulation frequencies of 1 kHz and 4 kHz are applied), which is limited by the used microheater. For the tracking of the frequency shift of all comb modes, we propose a two-wavelength i-MZI, in which the frequency shifts of the two of the comb modes are tracked. Since the frequencies of the comb modes are highly correlated as expressed with the only two degrees of freedom ($f_{\rm ceo}$: carrier-envelope offset frequency and $f_{\rm rep}$: comb mode spacing), the frequency shift of all comb modes can be estimated from the measurement of the two comb modes. 
%The demonstrated rapid and large scanning of the soliton comb with the frequency tracking not only enables higher-depth resolution massively parallel FMCW LiDAR \cite{riemensberger2020massively} and faster comb-resolved spectroscopy \cite{kuse2020continuous}, but also could unlock the use of the soliton comb for FMcomb LiDAR \cite{kuse2019frequency}, high-resolution optical vector analyzer \cite{qing2019optical}, and comb-resolved dense imaging by a single comb\cite{hase2018scan, bao2019microresonator}.

\begin{figure}[!ht]
\centering
\fbox{\includegraphics[width=0.85\linewidth]{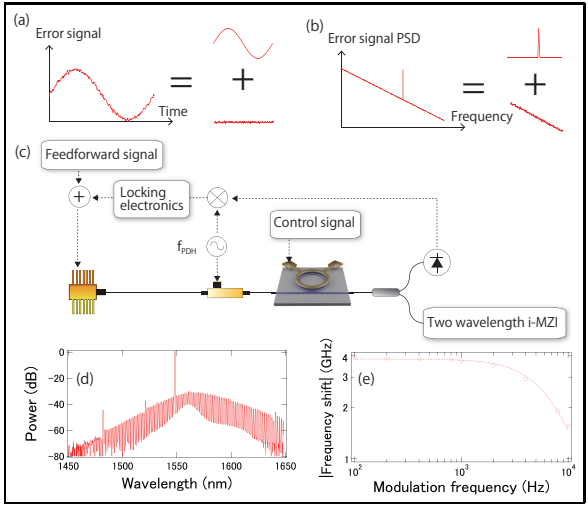}}
\caption{Illustrations of an error signal in the time domain (a) and frequency domain (b). (c) Schematic of the experimental setup. (d) The optical spectrum of the soliton comb. (e) The frequency response of the microheater when 10 mW electric power is applied. The dotted curve is a Lorentz function.}
\end{figure}

The basic concept of our scan system is illustrated in Figs. 1 (a) and (b). In our scan system with feedback and feedforward, a microheater deposited on a microresonator \cite{kuse2020continuous} is modulated by a pre-determined signal, which could be arbitrary (e.g. $A{\rm sin}(f_mt)$ in the demonstration, where $A$, $f_m$, and $t$ are modulation amplitude, modulation frequency, and time, respectively). In the time domain, an error signal is considered as the sum of a sine function and a random noise. In the frequency domain, a power spectral density (PSD) of the error signal is expressed as the sum of a single tone at $f_m$ and broadband noise. The sine function/single tone noise is generated from a control signal applied to the microheater. The random/broadband noise is originated from the noise of the pump cw laser and microresonator. The sine function (single tone) in the time domain (in the frequency domain) is suppressed by a feedforward signal, while the random (broadband) noise is suppressed by a feedback signal. 

An experimental setup is shown in Fig. 1 (c) (more detail is shown in the supplementary material). A microresonator based on Si$_3$N$_4$ (SiN, width = 2 $\mu$m and height = 800 nm) is used for the generation of a single soliton comb. The free spectral range (FSR) and loaded Q of the microresonator are about 192 GHz and $10^6$, respectively. A distributed feedback (DFB) laser around 1548 nm (AA1406 series from G\&H) is directed through a phase modulator (MPZ-LN-10 from iXblue), which is driven by an RF oscillator with the modulation frequency ($f_{\rm PDH}$). $f_{\rm PDH}$ determines the detuning between the DFB laser and the resonance frequency. The generated soliton comb is guided to a two-wavelength i-MZI to characterize the frequency shift of all comb modes. The details are explained later. The residual pump is photodetected to generate an error signal for PDH locking. When the frequency of the upper sideband that is generated from the phase modulator crosses the resonance frequency, the error signal, which is generated after demodulation by a double balanced mixer, has a zero-crossing voltage \cite{kuse2020continuous}. A feedforward (FF) signal is added to the feedback (FB) signal, making a ‘feedforward-feedback (FF-FB)’ signal. The feedforward-feedback (FF-FB) signal is then applied to the DFB laser. The optical spectrum of the single soliton comb used in the experiments is shown in Fig. 1(d). The comb mode spacing corresponds to the FSR of the microresonator. Figure 1(e) shows the response of the microheater when about 10 mW electric power is applied. The 3 dB modulation bandwidth of the microheater is about 7 kHz. 

\begin{figure}[!ht]
\centering
\fbox{\includegraphics[width=0.8\linewidth]{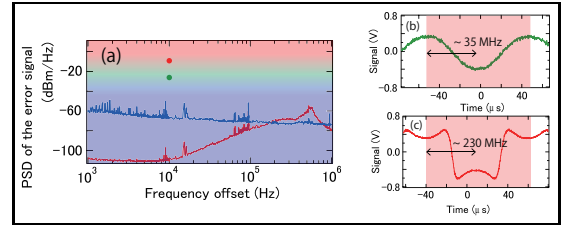}}
\caption{(a) PSD of the error signal for the free-running (blue) and locked (red) cases. The color map indicates how stable the detuning (i.e. soliton) is. See the definition in the main text. The green and red circles are the PSDs of the error signals at the modulation frequency when the DFB laser is modulated by 35 MHz for the green circle and 230 MHz for the red circle with the modulation frequency of 10 kHz, which corresponds to the error signal in the time domain (b) for the green circle and (c) for the red circle.}
\end{figure}

The blue curve in Fig. 2(a) shows the power spectral density (PSD) of the error signal when the detuning is slowly (< 100 Hz feedback bandwidth) locked at 1100 MHz, which is the middle of the soliton-existing detuning (from 800 MHz to 1400 MHz), by the feedback loop without any FF signal. In Fig. 2(a), we distinguish the power of the single tone noise ($P_{\rm tone}$) against the soliton-existing detuning range ($\Delta_{\rm sol}$); blue, green, and red zones are termed as ‘quiet’ ($P_{\rm tone}$ < 5 \% of $\Delta_{\rm sol}$), ‘safe’ (5 \% of $\Delta_{\rm sol}$ < $P_{\rm tone}$ < 33 \% of $\Delta_{\rm sol}$), and ‘fragile’ (33 \% of $\Delta_{\rm sol}$ < $P_{\rm tone}$) in this letter, respectively. In the ‘fragile’ zone, the soliton comb can be lost. When a tiny 10 kHz modulation is added to the DFB laser in addition to the slow feedback signal, the error signal in the time domain shows the oscillation at 10 kHz (Fig. 2(b)). The excursion of the error signal is 35 MHz, which corresponds to the green circle in the frequency domain in Fig. 2(a). The applied tiny single tone brings the error signal from ‘quiet’ to ‘safe’. When a larger 10 kHz modulation is added to the DFB laser, the error signal shows the nonlinear behavior due to the deviation from the linear range of the PDH locking (Fig. 2(c)). Assuming a linear response of the error signal, the excursion of the error signal is 230 MHz, which corresponds to the red circle in the ‘fragile’ zone in Fig. 2(a). When a feedback loop with a large locking bandwidth (> 100 kHz) is applied, the PSD of the error signal is significantly suppressed as shown in the red curve in Fig. 2(a). At the 10 kHz frequency offset, the feedback gain is more than 40 dB, which can bring both green and red circles into the ‘quiet’ zone. However, when the microheater is modulated by more than 25 GHz at 10 kHz, which is even 40 dB higher than the red circle, the feedback loop cannot even bring the PSD back to the ‘safe’ zone. Therefore, the only feedback signal is not enough for rapid and large comb mode scanning.

To overcome the limitations of the feedback-only system, the FF signal is added to the FB signal, generating an FF-FB signal. A procedure to make the FF-FB is described in the supplementary material. 
%A FF signal ($f_{\rm FF}(t)$) is created as follows. When a control signal ($f_{\rm heater}(t)$) is applied to the microheater, a FF signal $f^{(0)}_{\rm FF}(t)$ is set to be $A^{(0)}f_{\rm heater}(t)$. Then, the feedforward signal is updated as $f^{(1)}_{\rm FF}(t) = f^{(0)}_{\rm FF}(t) + Cf^{(0)}_{\rm FB}(t)$, where $f^{(0)}_{\rm FB}(t)$ is the FB signal. Another repetitive step as $f^{(n)}_{\rm FF}(t) = f^{(n-1)}_{\rm FF}(t) + Cf^{(n-1)}_{\rm FB}(t)$ is employed when $f^{(n-1)}_{\rm FF}(t)$ is not enough to reach the limit of the microheater (more data is shown in the supplementary material). 
Figure 3(a) shows an error signal when the microheater is modulated by about 2.3 GHz at 10 kHz. Because of the limited feedback gain, the error signal shows an oscillation (the blue curve in Fig. 3(a)), being in the ‘safe’ zone for the PSD of the error signal. When a feedforward is turned on, the oscillation is suppressed as shown in the red curve in Fig. 3(a). The PSD of the error signal at 10 kHz frequency offset is suppressed by more than 40 dB as shown in Fig. 3(b), being brought into the ‘quiet’ zone. In total, our control system has more than as large as 80 dB (40 dB by feedback and 40 dB by feedforward) gain at 10 kHz, which is large enough to reach the limit of the scan range of the microheater.

\begin{figure}[!ht]
\centering
\fbox{\includegraphics[width=\linewidth]{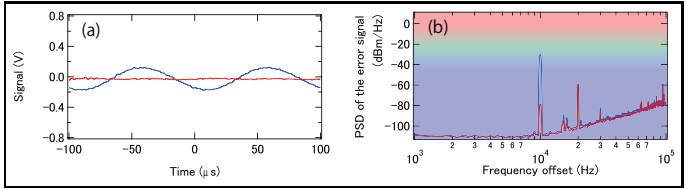}}
\caption{(a) Error signals with only feedback (blue curve) or both feedback and feedforward (red curve) when the microheater is modulated by 2.3 GHz at 10 kHz. (b) PSD of the error signal, which corresponds to (a).}
\end{figure}

With the proposed control system, the comb modes are scanned rapidly and largely. In the demonstration, the modulation frequencies of 1 kHz, 4 kHz, and 10 kHz with about 170 mW electric power are applied to the microheater. Although more electric power could be applied, we have not tried such high electric power not to break the microheater through electromigration. The frequency of the DFB laser during the comb mode scanning is measured by an i-MZI as shown in Fig. 4. The frequency shift is about 70 GHz, 52 GHz, and 26 GHz for 1 kHz (green curve in Fig. 4), 4 kHz (blue curve in Fig. 4), and 10 kHz (red curve in Fig. 4) modulation frequencies, which corresponds to the scan speed of more than 550 THz/s at the fastest. In the fastest case, the demonstrated scan speed is 1000 times faster than the previous demonstration \cite{lin2020broadband}. The scan range could be further improved by overcoming the scan range of our microheater, because the DFB laser can be also scanned more than 80 GHz at 10 kHz. A new microheater has a scan range of 210 GHz, 180 GHz, and 120 GHz at 1 kHz, 4 kHz, and 10 kHz, respectively (more data is shown in the supplementary material), although a single soliton comb cannot be generated due to a failure of the design of microresonators. 80 dB gain of the proposed method is large enough for the scanning of 80 GHz at 10 kHz, which is the limit of the DFB laser, to be in the ‘safe’ zone in the PSD of the error signal. To cover one FSR of soliton combs, microresonators with smaller FSR would be required.

\begin{figure}[!ht]
\centering
\fbox{\includegraphics[width=0.8\linewidth]{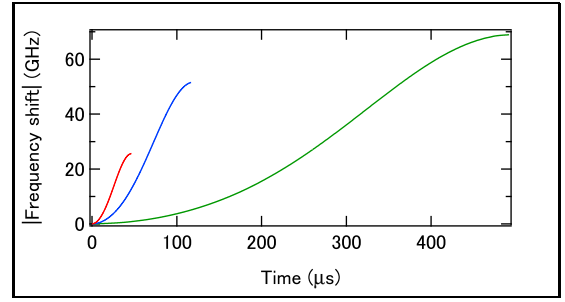}}
\caption{Frequency shift of a comb mode at 1 kHz (green), 4 kHz (blue), and 10 kHz (red) when the electric power of 170 mW is applied to the microheater.}
\end{figure}

The frequency shifts show the nonlinearity (close to a sine function in our experiment), which causes the degradation of the depth resolution and precision of LiDAR with the frequency-scanned soliton comb. Also, the frequency shifts should be precisely characterized when the frequency-scanned soliton comb is applied to spectroscopy. Therefore, a method to measure the frequency shift of all comb modes is proposed. Once the frequency shift of all comb modes is measured, the nonlinearity is corrected by signal processing \cite{kuse2019frequency}. The system consists of a two-wavelength i-MZI (the time delay between two arms is 1.35 ns) (Fig. 5(a)). Two outputs from the i-MZI are optically bandpass filtered to extract a single comb mode at different frequencies (e.g. $\nu_m$ and $\nu_n$), followed by photodetectors. The frequency shift of the m-th and n-th ($\delta_m$ and $\delta_n$) comb modes is estimated from the phase of the photodetected signals, which are expressed by using the frequency shift of the comb mode spacing and the carrier-envelope offset frequency ($\delta_{\rm rep}$ and $\delta_{\rm ceo}$) as
\begin{equation}
\delta_{m (n)} = m (n)\delta_{\rm rep} + \delta_{\rm ceo}.
\end{equation}
With the measurement of $\delta_m$ and $\delta_n$, the frequency shift of the k-th comb mode ($\delta_k$, k is an arbitrary integer) can be estimated because the comb modes have only two degrees of freedom (i.e. $f_{\rm rep}$ and $f_{\rm ceo}$). 
\begin{equation}
\delta_{k} = \delta_{n} + \frac{k - n}{n - m}\cdot \left(\delta_n - \delta_m\right).
\end{equation}
Figure 5(b) shows the frequency shift of the -12th (red curve in Fig. 5(b)) and +12th (blue curve in Fig. 5(b)) comb modes. The amount of the frequency shift between the two comb modes is different, indicating the comb mode spacing is changed when the microheater is used for the scan. $\delta_{\rm rep}$ can be also estimated from the measurement as 
\begin{equation}
\delta_{\rm rep} = \frac{\delta_{m} - \delta_{n}}{m -n}.
\end{equation}
The estimated $\delta_{\rm rep}$ is compared with the measurement, in which the frequency of a beatnote between the electro-optically generated sidebands from the neighboring comb modes is counted \cite{xue2016thermal} (more detail is shown in the supplementary material) by an RF spectrum analyzer. As shown in Fig. 5 (c), the two independent measurements of $\delta_{\rm rep}$ are well overlapped. Due to the limited data acquisition time of the used RF spectrum analyzer, the scan speed is 0.5 Hz in these measurements. To investigate the precision of the proposed measurement, another output is added to the i-MZI. The output with another optical bandpass filter and photodetector is used to measure the frequency shift of the -6th comb mode. The measured frequency shift of the -6th comb mode is compared with that estimated from the -12th and +12th comb modes. As shown in Fig. 5(d), The frequency difference between the measurement and the estimation is about 10 MHz over the frequency shift of 50 GHz. As shown in the supplementary material, even when the frequency shift of the same comb mode is measured, the estimated frequency shift shows a similar frequency error, which indicates the frequency error does not depend on the integer k as long as the k-th comb mode is between the m-th and n-th comb mode. On the contrary, when the integer k is outside of the measured two comb modes, the frequency error scales up with the factor of $\frac{k-n}{n-m}$. Also, when the same measurement is implemented without the soliton comb, the frequency error is smaller. Therefore, the frequency error is likely to be caused by the slowly-varying amplitude noise of the photodetected signal, inducing the phase error in our signal processing. The frequency error might be reduced by using an optical 90-degree hybrid coupler and balanced photodetection. 

\begin{figure}[!ht]
\centering
\fbox{\includegraphics[width=0.9\linewidth]{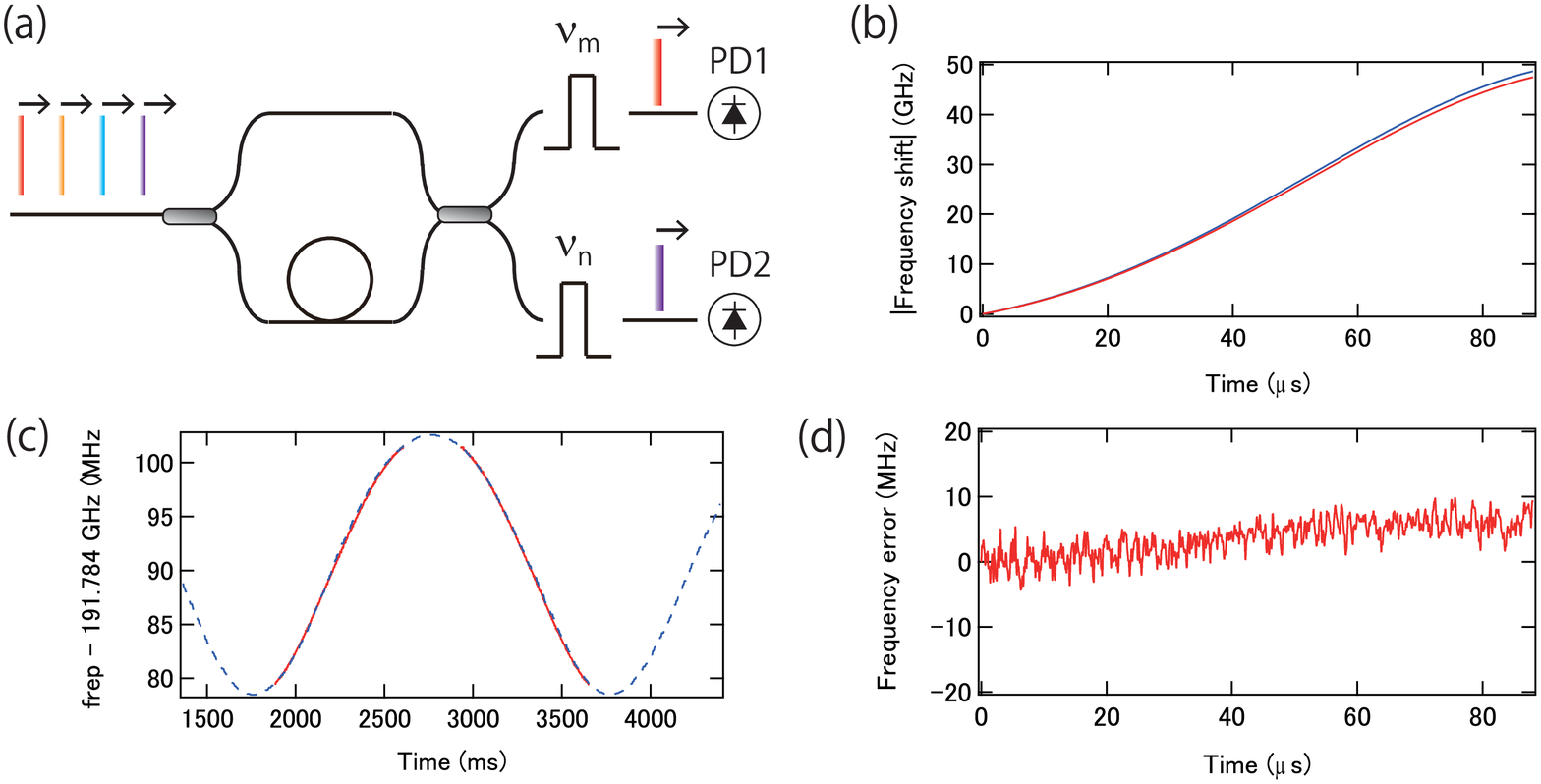}}
\caption{(a) Schematic of the two-wavelength i-MZI. (b) Frequency shift of the +12th (blue) and -12th (red) comb modes. (c) $f_{\rm rep}$ shift measured by the EO sideband method (dashed blue) and the two-wavelength i-MZI (red). (d) Frequency error of the two-wavelength i-MZI estimated from (b).}
\end{figure}

In the demonstration, although the only frequency shift is measured, the absolute frequency of all comb modes can be also estimated by referencing two known frequencies such as molecular absorption lines or two reference cw lasers whose absolute frequencies are known (more information in the supplementary material). To minimize the i-MZI, the fiber-based i-MZI can be replaced by optical waveguides on a chip, including optical couplers, delay, arrayed-waveguide gratings (or microrings) for optical bandpass filtering, and photodetectors. 

To summarize, we have developed the two fundamental techniques for the scanning of the soliton comb; (1) rapid and large scanning of the comb modes, and (2) measurement of the frequency shift of all comb modes. By successfully incorporating the feedback with the feedforward, a control gain beyond the feedback system has been obtained. With the proposed control method with 80 dB gain (more than 40 dB each from the feedback and feedforward) at the 10 kHz frequency offset, the comb mode scanning of 26 GHz at 10 kHz (70 GHz for 1 kHz, 52 GHz for 4 kHz) limited by the used microheater was obtained, which is equal to the scan speed of 550 THz/s. Also, the two-wavelength i-MZI has been proposed and demonstrated to measure the frequency shift of all comb modes of the soliton comb. The estimated frequency error was below 10 MHz over 50 GHz comb mode scanning. Given that 100 comb modes with the comb mode spacing of 50 GHz are scanned by 50 GHz at a 10 kHz rate, the soliton comb can cover 5 THz in 50 $\mu s$, which corresponds to a cw laser with the scan range and speed of 5 THz and 100 PHz/s, respectively. Therefore, the rapidly and largely scanned soliton comb in conjunction with a WDM system extremely accelerates the spectroscopic system using a frequency-scanned cw laser. Not only to replace the frequency-scanned cw laser, but the demonstrated frequency-scanned soliton comb could be readily applied to FMcomb LiDAR \cite{kuse2020continuous}, high-resolution optical vector analyzer \cite{qing2019optical}, and comb-resolved dense imaging \cite{hase2018scan, bao2019microresonator} in addition to higher-depth resolution massively parallel FMCW LIDAR \cite{riemensberger2020massively} and faster comb-resolved spectroscopy \cite{kuse2020continuous}.

\medskip
\noindent\textbf{Funding.}
This work was financially
supported by JST PRESTO (JPMJPR1905), Research Foundation for Opto-Science and Technology, KDDI Foundation, and Cabinet Office, Government of Japan (Subsidy for Reg. Univ. and Reg. Ind. Creation).

\medskip
\noindent\textbf{Disclosures.} The authors declare no conflicts of interest.

\medskip
\noindent\textbf{Supplemental document} See Supplement 1 for supporting content.
% Bibliography
\bibliography{sample}

% Full bibliography added automatically for Optics Letters submissions; the following line will simply be ignored if submitting to other journals.
% Note that this extra page will not count against page length
\bibliographyfullrefs{sample}

%Manual citation list
%\begin{thebibliography}{1}
%\bibitem{Zhang:14}
%Y.~Zhang, S.~Qiao, L.~Sun, Q.~W. Shi, W.~Huang, %L.~Li, and Z.~Yang,
 % \enquote{Photoinduced active terahertz metamaterials with nanostructured
  %vanadium dioxide film deposited by sol-gel method,} Opt. Express \textbf{22},
  %11070--11078 (2014).
%\end{thebibliography}

% Please include bios and photos of all authors for aop articles
\ifthenelse{\equal{\journalref}{aop}}{%
\section*{Author Biographies}
\begingroup
\setlength\intextsep{0pt}
\begin{minipage}[t][6.3cm][t]{1.0\textwidth} % Adjust height [6.3cm] as required for separation of bio photos.
  \begin{wrapfigure}{L}{0.25\textwidth}
    \includegraphics[width=0.25\textwidth]{john_smith.eps}
  \end{wrapfigure}
  \noindent
  {\bfseries John Smith} received his BSc (Mathematics) in 2000 from The University of Maryland. His research interests include lasers and optics.
\end{minipage}
\begin{minipage}{1.0\textwidth}
  \begin{wrapfigure}{L}{0.25\textwidth}
    \includegraphics[width=0.25\textwidth]{alice_smith.eps}
  \end{wrapfigure}
  \noindent
  {\bfseries Alice Smith} also received her BSc (Mathematics) in 2000 from The University of Maryland. Her research interests also include lasers and optics.
\end{minipage}
\endgroup
}{}

\end{document}